\documentclass[doublecol,hyperref]{epl2}
\usepackage{epsfig,color,amsfonts}
\usepackage{xcolor}
\usepackage{ulem}
\usepackage{bm} 
\usepackage{amsmath}
\usepackage{ulem}
 
\title{Martingale theory for housekeeping heat}
\shorttitle{Martingale theory for housekeeping heat} 

\author{Rapha\"el Ch\'etrite\inst{1} \and Shamik Gupta\inst{2} \and Izaak Neri\inst{3,4} \and \'Edgar Rold\'an\inst{5}}
\shortauthor{R. Ch\'etrite, S. Gupta, I. Neri and \'E. Rold\'an}

\institute{
\inst{1} Universit\'e C\^ote d'Azur, CNRS, LJAD, Parc Valrose, 06108
Nice Cedex 02, France \\
\inst{2} Department of Physics, Ramakrishna Mission Vivekananda
University, Belur Math, Howrah 711 202, India \\
\inst{3} Department of Mathematics, King's College London, Strand, London WC2R 2LS, United Kingdom\\
\inst{4} Max Planck Institute for the Physics of Complex Systems, N\"othnitzer Strasse 38, 01187 Dresden, Germany\\
\inst{5} ICTP - The Abdus Salam International Centre for Theoretical Physics, Strada Costiera 11, 34151 Trieste, Italy}


\abstract{
  The housekeeping heat is the energy exchanged between a system and its environment in a nonequilibrium process that results from the violation of detailed balance.
  We describe  fluctuations of the housekeeping heat in mesoscopic systems using the theory of martingales, a mathematical framework widely used in probability theory and finance. We show that the exponentiated housekeeping heat (in units of $k_{\rm B}T$,  with $k_B$ the Boltzmann constant and  $T$ the temperature) of a Markovian nonequilibrium process under arbitrary time-dependent driving is a martingale process. From this result, we derive universal equalities and inequalities for the statistics of stopping-times and suprema of the  housekeeping heat.  We test our results with numerical simulations of a system driven out of equilibrium and described by Langevin dynamics. 
}
\begin{document}
\maketitle

\section{Introduction and main results}
A real-valued stochastic process $M_t$ is a \textit{martingale}~\cite{Ville:39,Doob,Williams,LS} 
if it satisfies the following two conditions: (i) the expected value  of $M_t$
conditioned on its past history $M_{[0,\tau]}$ is 
\begin{equation}
\Big\langle M_t \,\Big|\, M_{[0,\tau]} \Big\rangle  = M_{\tau}, 
\label{eq:1}
\end{equation}
for any  time $\tau \leq t$, and (ii) the process is integrable, $\langle
|M_t|\rangle <\infty$ at all times $t$.  We use the notation $\langle X | Y \rangle
\equiv \int X P(X|Y)\mathrm{d}X$ for the conditional expectation of  $X$ given $Y$, and we write $P(X|Y)$ for the conditional probability density of $X$ given $Y$.  
 
We also consider martingales $M_t$ relative to a
stochastic process $X_t$.    Such martingale processes satisfy the following three conditions: (i) the expected valu of $M_t$ conditioned on the past history $X_{[0,\tau]}$ equals 
\begin{equation}
\Big\langle M_t \,\Big|\, X_{[0,\tau]} \Big\rangle  = M_{\tau}, 
\label{eq:1T}
\end{equation}
for any  time $\tau \leq t$,  (ii) the process is integrable $\langle |M_t|\rangle <\infty$ at all times $t\geq 0$, and (iii) $M_t = f(X_{[0,t]}, t)$ with $f$ a real-valued function.   Note that  (\ref{eq:1T}) is a generalization of (\ref{eq:1}). 

Examples of martingales are: (i) the Wiener process $\mathcal{W}_t$ (Brownian motion); (ii) a geometric Brownian motion $\Sigma_t = e^{\sigma \mathcal{W}_t - (1/2)\sigma^2 t}$, where  $\sigma $ is a real number; (iii) It\^{o} processes $I_t = \int_0^t F(X_{s})\cdot \mathrm{d} \mathcal{W}_s$, where $F(x)$ is  a real-valued and well-behaved function~\cite{Oks, Note}  and $\cdot$ denotes the It\^{o} product; (iv)  ratios of probability densities of trajectories $\Lambda_t=\mathcal{P}(X_{[0,t]})/\mathcal{Q}(X_{[0,t]})$ \cite{radon} are martingales relative to the process $X_t$ if the statistics of the process $X_t$ is generated by  $\mathcal{Q}$.  The
 average in~(\ref{eq:1T}) is then done with respect to the probability
 $\mathcal{Q}$, i.e., $\langle \Lambda_t \,|\, X_{[0, {\tau}]} \rangle_{\mathcal{Q}}  = \Lambda_\tau$, for all $t\geq \tau $,  where $\langle\cdot \rangle_\mathcal{Q}$  denotes expectation with respect to  $\mathcal{Q}$.   

Martingales have found widespread applications in mathematics, economics
and gambling~\cite{Doob,Williams,LS,Oks,Pliska}. Gambling strategies
can be modelled with \textit{stopping times} $\mathcal{T}$, which denote the time
when a gambler decides to cash out. Stopping times are random times when a stochastic process satisfies for the first time a certain criterion, which is non-anticipative. 
Examples of stopping times are first-passage times, second-passage times, etc.
Stopping times  depend on \textit{non-anticipative}
stopping rules 
i.e.  the stopping time $\mathcal{T}$ conditioned on $X_{[0,\mathcal{T}]}$ is independent of 
 the trajectory in the future  $X_{(\mathcal{T},\infty)}$.   An important result is Doob's optional stopping theorem~\cite{Doob}, which states there exist no winning gambling strategies that can make profit out of a martingale process,  irrespective of the stopping rule used:      
\begin{eqnarray}
\langle M_\mathcal{T}\rangle = \langle M_0\rangle, \label{eq:stopping}
\end{eqnarray}
i.e., the expected value of a martingale at a stopping time $\mathcal{T}$ equals the expected value at the initial time $t=0$. 
Doob's optional stopping theorem (\ref{eq:stopping}) holds under some
additional conditions on either the  stopping time or the martingale
process.  For example, (\ref{eq:stopping}) holds if the stopping time
$\mathcal{T}$ satisfies $\mathcal{T}\in[0,\tau]$ with $\tau$ a constant
fixed time, or, if $M_t$ is uniformly integrable.   Another important example 
for which (\ref{eq:stopping})  holds is when $\mathcal{T}$ is a
first-passage time with two absorbing thresholds $m_+ > M_0$ and
$m_-<M_0$ and if additionally $\textsf{Pr}(\mathcal{T}<\infty) = 1$,
i.e., the probability to quit the game after a finite time is one.   We  use the latter formulation of the stopping-time theorem in this paper.  

In physics,  martingales have  not been exploited much so
far~\cite{Neri:17,Chetrite:11,Ventejou:18, adler, bernard}. Recent work
has applied martingale theory to the thermodynamics of stochastic
processes~\cite{Neri:17,Chetrite:11}.   In particular,  it was found
that the exponential of the negative entropy production  of a stationary
stochastic process is a martingale.  Reference~\cite{Neri:17} uses this
observation to  derive fluctuation relation for stopping times of
entropy production and to derive universal laws on the statistics of
infima of the entropy production.   This so-called infimum law  was
tested experimentally in a double electronic dot~\cite{Singh:17}.
Martingales have also been discovered in the context of classical
quenched systems~\cite{Ventejou:18} and quantum
mechanics~\cite{adler,bernard}.  An interesting question is whether
there exist other physical processes, in particular, in the context of
stochastic energetics, which are martingales.

In this paper, we find a  martingale process that describes
fluctuations of the heat exchanged between a mesoscopic system and its
environment in a nonequilibrium process that takes place at isothermal
conditions.    Before presenting this martingale process, we review some
aspects of nonequilibrium thermodynamics. Following Oono and Paniconi \cite{Oono:98}, the fluctuating dissipated heat $ Q_t $ 
during a single realization of a nonequilibrium process can be
decomposed as the sum of two terms, a housekeeping heat  $ Q^{\rm hk}_t
$ and an excess heat $ Q^{\rm ex}_t $~\cite{Seki, Sasa}.  For processes
that relax to an equilibrium state, one has $Q^{\rm hk}_t
 = 0$, whereas for nonequilibrium steady states, $\langle Q^{\rm
ex}_t \rangle = 0$.  Moreover, the housekeeping heat is on average
negative, i.e., $\langle  Q^{\rm hk}_t\rangle\leq 0$ for all times $t$;
in other words, a stochastic process has on average a tendency to
dissipate housekeeping heat.   Note that here we use the convention that
all heat exchanges are negative when heat flows from the system to its
environment. We will also  assume that the environment is isothermal at temperature~$T$. 

Fluctuations of the housekeeping heat also obey universal laws.  The
housekeeping heat satisfies an integral fluctuation relation: $\langle
e^{\beta Q^{\rm hk}_t} \rangle = 1$  \cite{Sasa, Speck:05, Esposito,
VDB}. Here, we  use the notation $\beta = (k_{\rm B}T)^{-1}$, with
$k_{\rm B}$ the Boltzmann constant.       
The integral fluctuation relation implies that fluctuations of the
housekeeping heat away from the average tendency to dissipate are rare.
Indeed, using the Markov inequality on  $e^{\beta Q^{\rm hk}_t}$ and
 the   integral fluctuation relation, we find that
\begin{eqnarray}
\text{Pr}\left( Q^{\rm hk}_t \geq q \right) \leq e^{-\beta q}, \quad q\geq 0.
\label{boundhouse}
\end{eqnarray}
 Hence, the probability to observe a system that absorbs at least an
 amount of housekeeping heat $q$  is smaller than  $e^{-\beta q}$, so
 that large fluctuations away from the tendency of a nonequilibrium processes to dissipate occur rarely.   

In this paper, we show that the bound (\ref{boundhouse}) on the fluctuations of positive housekeeping heat can be significantly improved.   In particular, we  show that 
\begin{eqnarray}
\text{Pr}\left( {\rm sup}_{\tau\in[0,t]} Q^{\rm hk}_\tau \geq q \right) \leq e^{-\beta q}, \quad q\geq 0,
\label{boundhousex}
\end{eqnarray}   
where ${\rm sup}_{\tau\in[0,t]} Q^{\rm hk}_\tau$ is the supremum of the dissipated heat over the full trajectory in the time interval  $[0,\tau]$.  
 The bound (\ref{boundhousex}) is tighter than the bound
 (\ref{boundhouse}), $\text{Pr}\left( Q^{\rm hk}_t \geq q \right) \leq \text{Pr}( {\rm sup}_{\tau\in[0,t]} Q^{\rm hk}_\tau \geq q ) \leq e^{-\beta q} $, since the supremum of a trajectory is always larger
 than its final value, i.e., ${\rm sup}_{\tau\in[0,t]} Q^{\rm hk}_\tau \geq
 Q^{\rm hk}_{t}$.   Remarkably, we show that for  continuous stochastic processes, the inequality (\ref{boundhousex})  becomes an equality.
 Hence,  the probability to observe a maximal fluctuation of the
 housekeeping heat larger than $q$ approaches a time-independent limit equal to $e^{-\beta q}$, which is linked to the fact that $Q^\text{hk}_t$ has an average tendency to decrease in time. 
 We also derive an integral fluctuation relation at stopping times:
      \begin{equation}
 \Big\langle e^{\beta Q^{\rm hk}_\mathcal{T}} \Big\rangle = 1. \label{eq:intStopp_intro}
 \end{equation}  
 We remark here that in contrast to the standard fluctuation relation
 $\langle e^{\beta Q^{\rm hk}_t} \rangle=1$~\cite{Sasa, Speck:05,
 Esposito, VDB}, the average in~(\ref{eq:intStopp_intro}) is over
 trajectories of \textit{different} time duration, since $\mathcal{T}$
 is a random time. This  fluctuation relation~\eqref{eq:intStopp_intro} holds for a broad class of stopping times $\mathcal{T}$, including first-passage times that are bounded~\cite{bounded} and first-passage times with two absorbing boundaries. Using Jensen's inequality and Eq.~(\ref{eq:intStopp_intro})
 gives 
 \begin{equation}
 \Big\langle Q^{\rm hk}_\mathcal{T} \Big\rangle \leq 0,   \label{Qhk}
 \end{equation}
 which implies that it is not possible to  extract housekeeping heat  from the thermal reservoir, irrespective of the stopping protocol $\mathcal{T}$  used to end the stochastic process $X_t$.   In other words, whatever stopping strategy we use it is not possible that on average a system absorbs housekeeping heat from its environment.
 

 The key insight to prove relations
 (\ref{boundhousex}-\ref{eq:intStopp_intro}) is that the process
 $e^{\beta Q^{\rm hk}_t}$ is a martingale with respect to the physical
 process $X_t$ that dissipates $Q^{\rm hk}_t$, i.e., for any $t\geq
 \tau$, 
 \begin{equation}
\Big\langle e^{\beta Q^{\rm hk}_{t}} \,\Big|\, X_{[0,\tau]} \Big\rangle  = e^{\beta Q^{\rm hk}_{\tau}}.
\end{equation}
This result is the cornerstone of this paper, and as we show below it holds for general Markov processes, which include Langevin processes and Markov jump processes. Note that these results share a similarity with those derived in \cite{Neri:17} for the stochastic entropy production, with the important distinction that the results here are also valid for non-stationary nonequilibrium processes.      

 \section{Nonequilibrium heat fluctuations of diffusions} Before discussing general Markov processes, we  consider  heat fluctuations in  $n$-dimensional
It\^{o}-diffusion processes of the form
\begin{equation}
\frac{\mathrm{d}\vec{X}_t}{\mathrm{d}t} =  \left[\vec{F}_t + \vec{\nabla}\cdot\mathbf{D}_t\right](\vec{X}_t) + \sqrt{2\mathbf{D}_t(\vec{X}_t)}\;\cdot\vec{\xi}_t,
\label{eq:langevin}
\end{equation}
where $\vec{F}_t $ is a $n$-dimensional vector field and $\mathbf{D}_t$ is an invertible diffusion matrix field of dimension $n\times n$.  The subindex $t$ denotes an explicit time dependence. 
The noise $\vec{\xi}_t$ is a $n$-dimensional Gaussian  white noise with zero mean and covariance matrix equal to the identity matrix.
We assume that the Einstein relation holds such that  $\vec{F}_t = \vec{f}_t -\beta\mathbf{D}_t  \vec{\nabla}U_t$ with $\vec{f}_t$ an external force
and $U_t$ a potential. The presence of the term
$\vec{\nabla}\cdot\mathbf{D}_t$ in \eqref{eq:langevin} ensures that in
the case without external force and with time-independent
potential, the stationary distribution is the
Boltzmann distribution with potential $U$ and inverse temperature $\beta$ \cite{Ku1,Ku2, Helvetica}. Note that we could
have considered other conventions for the stochastic discretization in~(\ref{eq:langevin}),
e.g., the Fisk-Stratonovich convention~\cite{fisk,stratonovich}, but the martingality of a process is revealed in a simple mathematical form when using It\^{o} convention. 
 The  Fokker-Planck equation associated with the process   (\ref{eq:langevin}) is
\begin{equation}
\partial_t P_t = L_t^{\dagger}P_t,
\label{eq:fpe}
\end{equation}
with
\begin{equation}
L_{t}=  \vec{F}_t\cdot \vec{\nabla}+\vec{\nabla}\cdot \mathbf{D}_t\cdot \vec{\nabla}
\label{eq:L}
\end{equation}
its Markovian generator. In~\eqref{eq:fpe}, $ L_t^{\dagger} = -\vec{\nabla}\cdot  \vec{F}_t +\vec{\nabla}\cdot \mathbf{D}_t\cdot \vec{\nabla}$ is the formal adjoint of the generator. We define the \textit{accompanying distribution}~\cite{hanggi} $\pi_t$ associated to the dynamics  as
\begin{equation}
L^{\dagger}_t \pi_t =0.
\label{eq:Ac}
\end{equation}
 We remark that  $\partial_t \pi_t \neq L_t^{\dagger}\pi_t$, i.e.,
 $\pi_t$ does not obey the Fokker-Planck equation because the right-hand
 side of this equation is zero whereas $\partial_t \pi_t \neq 0$.
 However, the accompanying distribution would be the stationary
 distribution for a process with the value of the external parameters constant and equal to  
those at a given time $s$ (i.e., a process with a generator $L_t = L_s$ for any $t \geq s$).   Notice that when the system is driven quasistatically one has $P_t=\pi_t$.

We now review the stochastic thermodynamics associated with the nonequilibrium fluctuations of It\^{o} processes~(\ref{eq:langevin}).    Such systems possess two type of forces that drives them out of equilibrium: (i) the non-conservative external force $\vec{f}_t$, and 
(ii) the explicit time dependence of the force and diffusion fields.
These two sources that drive the system out of equilibrium have distinct thermodynamic fingerprints in terms of
heat fluctuations, namely,   (i) the fluctuating  housekeeping
heat~\cite{Oono:98,Sasa}, which for the diffusion
process~(\ref{eq:langevin}) equals
\begin{equation}
Q^{\rm hk}_t\equiv k_{\rm B} T\int_{0}^{t}\left[\vec{\nabla}\ln\pi_{s}-\mathbf{D}_{s}^{-1} \cdot \vec{F}_{s}\right]\!(\vec{X}_{s})\circ\mathrm{d}\vec{X}_{s},
\label{eq:qhk}
\end{equation}\\
and (ii) the fluctuating excess heat~\cite{Oono:98,Sasa} 
\begin{equation}
Q^{\rm ex}_t\equiv - k_{\rm B}T\int_{0}^{t}\left[ \vec{\nabla}\ln\pi_{s}\right]\! (\vec{X}_{s})\circ \mathrm{d}\vec{X}_{s},
\label{eq:qex}
\end{equation} 
where $\circ$ denotes both the use of the Fisk-Stratonovich convention~\cite{fisk,stratonovich} and the scalar product. Using Stratonovich calculus, we can write the latter expression as
\begin{equation}
Q^{\rm ex}_t =- k_{\rm B}T\ln \frac{\pi_t (\vec{X}_t)}{\pi_0(\vec{X}_0)} + k_{\rm B}T \int_{0}^{t} \text{d}s \left[\partial_s\ln\pi_{s}\right]\!(\vec{X}_{s}). \label{eq:qex2}
\end{equation}
Summing (\ref{eq:qhk}) and (\ref{eq:qex}), we find that the total
fluctuating heat associated with a stochastic trajectory~\cite{Seki},
\begin{equation}
Q_t \equiv Q^{\rm hk}_t + Q^{\rm ex}_t,
\label{eq:dec}
\end{equation}
is equal to Sekimoto's expression~\cite{Seki} 
\begin{equation}
Q_t = - k_{\rm B}T\int_0^t   \vec{F}_s (\vec{X}_s) \cdot \mathbf{D}^{-1}_s(\vec{X}_s)  \circ \mathrm{d}\vec{X}_s.
\end{equation}

Equations~(\ref{eq:Ac}) and (\ref{eq:qhk})  imply that the
housekeeping heat vanishes in the absence of  external forces, i.e., when
$\vec{f}_t=\vec{0}$.   Nevertheless,  heat is dissipated through the excess heat which becomes a border term
$Q^{\rm ex}_t =- k_{\rm B}T\ln (\pi_t (X_t)/\pi_0(X_0))$, see~(\ref{eq:qex2}).  Therefore, the 
(asymptotic) rate of  $Q^{\rm ex}_t$ vanishes  when there is no explicit
time dependence in the dynamics.   In a steady
state process without explicit time dependence in the dynamics the housekeeping heat is non-zero, and this is the
reason behind the  name "housekeeping". An analogous formulation in terms of
adiabatic and non-adiabatic entropy production can be found in, e.g., Refs.~\cite{Esposito,VDB}.

\section{Martingale features of housekeeping heat for general Markov processes}    
In this paragraph we derive generic expression for the housekeeping heat of a Markov process, which is equal to (\ref{eq:qhk}) for  It\^{o}-diffusion processes.   This generic expression allows us to derive universal fluctuation properties of housekeeping heat.  
  
For a Markov process, the housekeeping heat associated with a trajectory $X_{[0,t]}$ of a general Markov process can be written as~\cite{CMP,Reinaldo:10,Harris}
\begin{equation}
 Q^{\rm hk}_t =k_{\rm B}T \ln\frac{\mathcal{P}^{\star}\left[X_{[0,t]}\right]}{\mathcal{P}\left[X_{[0,t]}\right]} .
 \label{eq:qhkstar}
\end{equation}
Here $\mathcal{P}[X_{[0,t]}]$ is the probability density of a trajectory $X_{[0,t]}$ with initial condition $\rho_0$ and the original dynamics given by $L_t$. On the other hand, $\mathcal{P}^{\star}[X_{[0,t]}]$ is the probability density of   the \textit{same} path $X_{[0,t]}$ with the same initial condition $\rho_0$ but under a different "$\star$ dynamics" (often called $\pi-$dual dynamics) given by
\begin{equation}
L_t^{\star} = \pi_t^{-1} L_t^{\dagger} \pi_t.
\label{eq:Lstar}
\end{equation}
Notably, for general Markov processes, $Q_t^{\rm hk}=0$ for all $t$ when the \textit{instantaneous} detailed balance condition $\pi_t^{-1} L_t^{\dagger} \pi_t = L_t$ is satisfied and thus $\mathcal{P}^{\star}=\mathcal{P}$. This clarifies the adjective "housekeeping"  from the fact that it corresponds to the heat absorbed by the system as a result of the violation of instantaneous detailed balance. 

Due to the fact that $e^{\beta Q^{\rm hk}_t}=\mathcal{P}^{\star}(X_{[0,t]})/\mathcal{P}(X_{[0,t]})$ is a path-probability ratio,  we obtain directly the cornerstone of this work: for general Markov nonequilibrium processes, the  exponentiated housekeeping heat in $k_{\rm B}T$ units  is a martingale relative to $X^t_0$, 
\begin{equation}
\Big\langle e^{\beta Q^{\rm hk}_{t}} \,\Big|\, X_{[0,\tau]} \Big\rangle  = e^{\beta Q^{\rm hk}_{\tau}},
\label{eq:1HK}
\end{equation}
for any $t\geq \tau$. Note that the average here is done with respect to $\mathcal{P}$. 
Notably, for $\tau=0$, $X_{[0,\tau]}=X_0$, and (\ref{eq:1HK})
implies the integral fluctuation relation~$\langle e^{\beta Q^{\rm
hk}_t}\rangle=1$~\cite{Speck:05}.  We remark that  this property holds
for generic Markov processes, which include diffusion processes~\cite{CMP} but also jump processes~\cite{Harris,esposito07}.

The martingality of $e^{\beta Q^{\rm hk}_t}$ entails second-law-like
relations for the housekeeping heat. From convexity of $-\ln,$  we find
that the housekeeping heat is a \textit{supermartingale} for general
Markov nonequilibrium processes. Its expected value in the future
$t\geq \tau$ is smaller than or equal to its value at time $\tau$:
\begin{equation}
\Big\langle  Q^{\rm hk}_t \,\Big|\, X_{[0,\tau]} \Big\rangle  \leq  Q^{\rm hk}_{\tau}.
\label{eq:hksupm}
\end{equation}
Note that for $\tau=0$, (\ref{eq:hksupm}) implies that $\langle
Q^{\rm hk}_t \,\rangle\leq 0$, as expected.
 
We   now show that (\ref{eq:qhkstar}) is consistent with expression (\ref{eq:qhk}) for the housekeeping heat in It\^{o}  diffusions.  
Note that in  the particular case of It\^{o} diffusions, given by (\ref{eq:langevin}), the "$\star$ dynamics" generates a stochastic process $\vec{Y}_t$ described by the Langevin equation
\begin{equation}
\frac{\mathrm{d}\vec{Y}_t}{\mathrm{d}t} \!=\!\!  \left[-\vec{F}_t +2\mathbf{D}_t\!\cdot\!\vec{\nabla} \ln \pi_t+  \vec{\nabla}\cdot\mathbf{D}_t\right]\! (\vec{Y}_t) + \sqrt{2\mathbf{D}_t(\vec{Y}_t)}\cdot\vec{\xi}_t. 
\end{equation}
We   use the It\^{o} convention, as in (\ref{eq:langevin}), to express  the housekeeping heat in (\ref{eq:qhk}):
\begin{eqnarray}
Q^{\rm hk}_t&=& k_{\rm B}T \int_{0}^{t}\left[\vec{\nabla}\ln\pi_{s}-\mathbf{D}_{s}^{-1} \cdot \vec{F}_{s}\right]\!(\vec{X}_{s})\cdot \mathrm{d}\vec{X}_{s} \label{eq:10} \\
&+& k_{\rm B}T \int_{0}^{t}\text{d}s \left[  \mathbf{D}_s\cdot  \vec{\nabla}\cdot \left(   \vec{\nabla}\ln\pi_{s}-\mathbf{D}_{s}^{-1} \cdot \vec{F}_{s} \right)  \right]\!(\vec{X}_s).\nonumber
\end{eqnarray}
From (\ref{eq:10}), we prove in the appendix that the exponentiated housekeeping heat satisfies the equation
\begin{equation}
\frac{\mathrm{d}}{\mathrm{d}t} e^{\beta Q^{\rm hk}_t} =  - \sqrt{2}  e^{\beta Q^{\rm hk}_t} \left[\frac{\vec{J}^{\pi}_t \cdot \mathbf{D}_t^{-1/2}}{\pi_t}\right]\!(\vec{X}_t) \cdot \vec{\xi}_t.  
\label{eq:15}
\end{equation}
Here, $\vec{J}_t^{\pi}\equiv \vec{F}_t \pi_t - \mathbf{D}_t\cdot
\vec{\nabla}\pi_t$ is the  effective current associated to $\pi_t$, and
$\vec{\xi}_t$ is the \textit{same} Gaussian white noise as in (\ref{eq:langevin}). Equation~\eqref{eq:15} reveals explicitly the
martingale structure of $e^{\beta Q^{\rm hk}_t}$: due to the absence of
drift terms on the right-hand side of~(\ref{eq:15}), $e^{\beta Q^{\rm hk}_t}$ is an It\^o process and therefore a martingale.
 An analogous equation  for the
exponentiated negative entropy production in nonequilibrium steady
states was reported in Ref.~\cite{Simone}. 
Equations of the type~(\ref{eq:15})  are also used in finance to model pricing options with stochastic  volatility~\cite{Heston}.

 For diffusion processes, the supermartingality of $Q^{\rm hk}_t$
is revealed by using It\^{o}'s lemma in~\eqref{eq:15}, leading to~\cite{Simone,Chun} 
\begin{equation}
\beta \frac{\text{d}Q^{\rm hk}_t}{\text{d}t}  =  - v^{\rm hk}(\vec{X}_t)
- \sqrt{2v^{\rm hk}(\vec{X}_t)}\cdot \xi^{\rm hk}_t(\vec{X}_t),
\label{eq:QhkL}
\end{equation}
with 
\begin{equation}
v^{\rm hk}(\vec{X}) \equiv  \left[\frac{\vec{J}^{\pi}_t \cdot \mathbf{D}_t^{-1} \cdot \vec{J}^{\pi}_t }{\pi_t^2}\right]\!(\vec{X})
\end{equation}
the "entropic drift"~\cite{Simone} of the housekeeping heat, and
\begin{equation}
\xi^{\rm hk}_t (\vec{X}) \equiv   \left[ \frac{  \mathbf{D}_t^{-2}\cdot\vec{J}^{\pi}_t}{\sqrt{\vec{J}^{\pi}_t\cdot \mathbf{D}^{-1}_t\cdot \vec{J}^{\pi}_t}}\cdot \vec{\xi}_t\right](\vec{X})
\end{equation}
a one-dimensional Gaussian white noise.
We remark here that the evolution of $Q^{\rm hk}_t$ given by (\ref{eq:QhkL}) is driven by the dynamics of $\vec{X}_t$, which is autonomous and described by~(\ref{eq:langevin}). Since $v^{\rm hk}(\vec{X})\geq 0$ for all values of $\vec{X}$, (\ref{eq:QhkL}) reveals that the housekeeping heat is a supermartingale. 

 \section{Universal stopping and extreme-value statistics from the martingale $e^{\beta Q^{\rm hk}_t}$}    
We  now discuss the physical implications of the martingale property of $e^{\beta Q_{\rm hk}}$ for the thermodynamics of  nonequilibrium processes. 
  First, we  apply  Doob's optional sampling theorem (\ref{eq:stopping}) to the
  martingale process $ e^{\beta Q^{\rm hk}_t}$, and find an integral fluctuation relation at stopping times~\cite{Neri:17}
    \begin{equation}
 \Big\langle e^{\beta Q^{\rm hk}_\mathcal{T}} \Big\rangle = 1. \label{eq:intStopp}
 \end{equation} 
 Applying Jensen's inequality $e^{\beta \langle Q^{\rm hk}_{\mathcal{T}}\rangle } \leq \langle e^{\beta  Q^{\rm hk}_{\mathcal{T}} } \rangle = 1$ we obtain $\langle Q^{\rm hk}_{\mathcal{T}} \rangle \leq 0$. Thus, a nonequilibrium system cannot on average
  absorb housekeeping heat from its environment by applying stopping rules based on measurements of $Q^{\rm hk}_t$.   The relation (\ref{eq:intStopp}) holds whenever the Doob's optional sampling theorem holds, for example, when $\mathcal{T}$ is a bounded stopping time~\cite{bounded} or when $e^{\beta Q^{\rm hk}_t}$ is uniformly integrable. 
 
The relation  (\ref{eq:intStopp}) also holds for two-boundary first-passage times $\mathcal{T}_{\rm FP}$ with thresholds $-q_-<0$ or  $q_+>0$: $\mathcal{T}_{\rm FP}$  is   the first-time when   $Q^{\rm hk}_t$ reaches either  $-q_-<0$ or  $q_+>0$.   Eq.~\eqref{eq:intStopp} implies then 
  \begin{equation}
 \Big\langle e^{\beta Q^{\rm hk}_{\mathcal{T}_{\rm FP}}} \Big\rangle = 1.    \label{eq:intStoppT}
 \end{equation}  
 Relation (\ref{eq:intStoppT}) allows us to derive several universal results about the fluctuations of $Q^{\rm hk}_{t}$.
 
A process that does not satisfy detailed balance dissipates heat and therefore
\begin{eqnarray}
\text{Pr}\left(\mathcal{T}_{\rm FP}<\infty\right) =  \mathsf{P}_- +
\mathsf{P}_ + = 1, \label{stopx}
 \end{eqnarray}
with 
\begin{equation}
\mathsf{P}_- \equiv \text{Pr}\left[Q^{\rm hk}_{\mathcal{T}_{\rm FP}} \leq -q_-\right], \; \mathsf{P}_+ \equiv \text{Pr}\left[Q^{\rm hk}_{\mathcal{T}_{\rm FP}} \geq q_+\right].
\end{equation}
The expression (\ref{eq:intStoppT}) can thus be written as 
\begin{eqnarray}
\mathsf{P}_- \langle e^{\beta Q^{\rm hk}_{\mathcal{T}_{\rm
FP}}}\rangle_{-} + \mathsf{P}_+  \langle e^{\beta Q^{\rm
hk}_{\mathcal{T}_{\rm FP}}}\rangle_{+} = 1, \label{eq:still}
\end{eqnarray} 
where $\langle e^{\beta Q^{\rm hk}_{\mathcal{T}_{\rm FP}}}\rangle_{-}$
and $\langle e^{\beta Q^{\rm hk}_{\mathcal{T}_{\rm FP}}}\rangle_{+}$ are
the conditional expectations of $e^{\beta Q^{\rm hk}_{\mathcal{T}_{\rm
FP}}}$ given that $Q^{\rm hk}_{\mathcal{T}_{\rm FP}} \leq -q_-$ or $Q^{\rm
hk}_{\mathcal{T}_{\rm FP}} \geq q_+$, respectively. Solving (\ref{stopx}) and
(\ref{eq:still}) gives
\begin{eqnarray}
\mathsf{P}_- = \frac{ \langle e^{\beta Q^{\rm hk}_{\mathcal{T}_{\rm FP}}}\rangle_{+}-1}{ \langle e^{\beta Q^{\rm hk}_{\mathcal{T}_{\rm FP}}}\rangle_{+}-\langle e^{\beta Q^{\rm hk}_{\mathcal{T}_{\rm FP}}}\rangle_{-}},\label{eq:hk1x}\\ \mathsf{P}_+ = \frac{1-\langle e^{\beta Q^{\rm hk}_{\mathcal{T}_{\rm FP}}}\rangle_{-}}{\langle e^{\beta Q^{\rm hk}_{\mathcal{T}_{\rm FP}}}\rangle_{+}-\langle e^{\beta Q^{\rm hk}_{\mathcal{T}_{\rm FP}}}\rangle_{-}}.\label{eq:hk2x}
\end{eqnarray} 

If the process $Q^{\rm hk}_t$ is {\it continuous}, then $Q^{\rm
hk}_{\mathcal{T}_{\rm FP}} = -q_-$ or $Q^{\rm hk}_{\mathcal{T}_{\rm FP}} = q_+$, and we find
exact expressions for the housekeeping heat splitting probabilities
$\mathsf{P}_-$ and $\mathsf{P}_+$: 
\begin{eqnarray}
\mathsf{P}_- = \frac{e^{\beta q_+}-1}{e^{\beta q_+}-e^{-\beta
q_-}},\quad\mathsf{P}_+ = \frac{1-e^{-\beta q_-}}{e^{\beta
q_+}-e^{-\beta q_-}}.\label{eq:firstCont}
\end{eqnarray} 
Note that the expressions (\ref{eq:firstCont}) hold for the It\^{o} diffusions  (\ref{eq:langevin}).
Consider now the  maximum value of the housekeeping heat: 
\begin{eqnarray}
Q^{\rm hk}_{\rm max} \equiv {\rm max}_{t\geq 0}Q^{\rm hk}_t,
\end{eqnarray}
i.e., the maximum amount of heat absorbed by the system, which
characterizes the largest fluctuation that violates the average tendency $\langle  Q^{\rm hk}_t\rangle\leq 0$
to dissipate.  The cumulative distribution of  $Q^{\rm hk}_{\rm max}$ is related to $\mathsf{P}_+$ by
\begin{eqnarray}
\text{Pr}\left[Q^{\rm hk}_{\rm max} \geq q_+ \right]=\lim_{q_-\rightarrow \infty}\mathsf{P}_+= e^{-\beta q_+}, \label{excatj}
\end{eqnarray}  
with $q_+ \geq 0$.
Hence, the probability density of $Q^{\rm hk}_{\rm max}$ is an exponential distribution  with mean value
\begin{eqnarray}
\langle Q^{\rm hk}_{\rm max} \rangle = k_BT. 
\end{eqnarray}  
If we consider the maximum value of $Q_{\rm hk}$ at a finite time: 
\begin{eqnarray}
Q^{\rm hk}_{\textrm{max},t} \equiv \max_{\tau \in [0,t]}\, Q^{\rm
hk}_\tau, 
\end{eqnarray}
then 
\begin{equation}
\text{Pr}\left[Q^{\rm hk}_{{\rm max},t} \geq q_+ \right] \leq
e^{-\beta q_+},\label{eq:lastnew}
\end{equation}
and 
\begin{equation}
 \langle Q^{\rm hk}_{{\rm max},t} \rangle \leq
k_BT, \label{eq:lastnew2}
\end{equation} 
since $Q^{\rm hk}_{\textrm{max},\tau} \leq Q^{\rm hk}_{\rm max}$. 

Although the relations (\ref{eq:firstCont}-\ref{eq:lastnew}) hold for continuous stochastic processes that do not satisfy detailed balance,  $e^{\beta Q^{\rm hk}_t}$ is  also a martingale for processes that may have jumps.   
 We therefore ask: what does the martingality of  $e^{\beta Q^{\rm hk}_t}$  imply for right-continuous stochastic processes, such as Markov jump processes?

\begin{figure}[ht!]
\centering
\includegraphics[width=.4\textwidth]{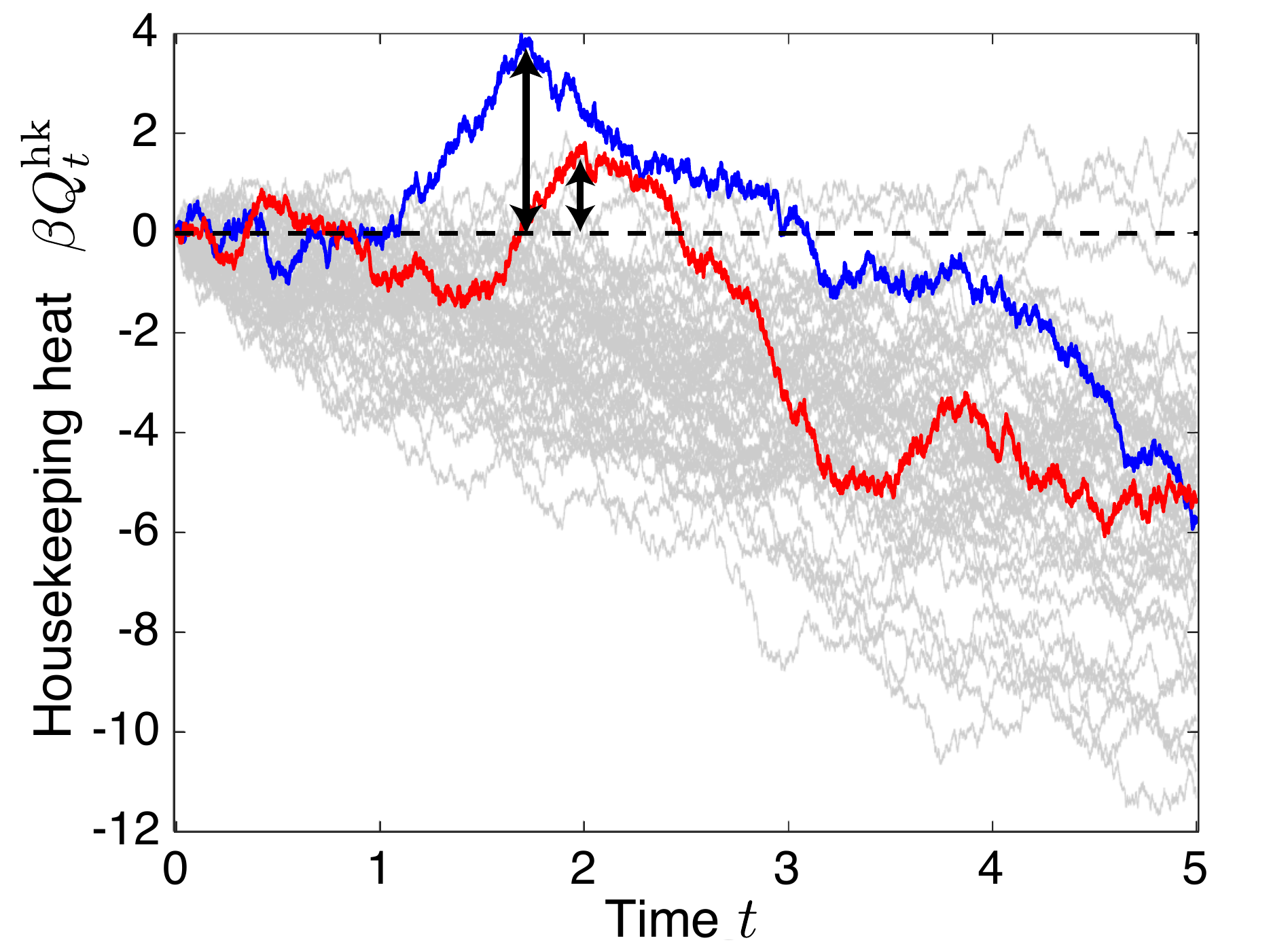}
\caption{Trajectories of  the fluctuating housekeeping heat $Q^{\rm
hk}_t$ (grey lines). The maximum value up to $t=5$, $Q^{\rm hk}_{\text{max},5},$ of two highlighted trajectories of the housekeeping heat (red and blue lines) 
 are indicated with double arrows. The 
time series are obtained from numerical simulations of the model~(\ref{eq:m1}), with $D=\omega=\beta=r=1.0$ with initial distribution $\rho_0(\theta)=\delta(\theta)$.
 \label{fig:traj}}
\end{figure}

  For {\it right-continuous} stochastic processes that do not satisfy detailed balance, the
  relations~(\ref{eq:hk1x}) and (\ref{eq:hk2x}) still hold.   We  use in
  them    
  \begin{equation}
 \langle e^{\beta Q^{\rm hk}_{\mathcal{T}_{\rm FP}}}\rangle_{-}  =
 e^{-\beta q_- } \langle e^{-\beta \Delta_-} \rangle_{-}, \, \langle
 e^{\beta Q^{\rm hk}_{\mathcal{T}_{\rm FP}}}\rangle_{+}  =  e^{\beta q_+
 } \langle e^{\beta \Delta_+} \rangle_{+},   
\end{equation}  
where $\Delta_-$ and $\Delta_+$  are the overshoot variables defined by
$-q_- -Q^{\rm hk}_{\mathcal{T}_{\rm FP}}$ and  $Q^{\rm
hk}_{\mathcal{T}_{\rm FP}}-q_+$ at the negative and positive threshold,
respectively.   Both   $\Delta_-$ and  $\Delta_+$   are positive random
variables, and for continuous processes they are equal to zero.    We
see that in the limit of $q_-\rightarrow \infty$,  $\langle e^{\beta
Q^{\rm hk}_{\mathcal{T}_{\rm FP}}}\rangle_{-}\rightarrow 0$, and
therefore, 
\begin{eqnarray}
\text{Pr}\left[Q^{\rm hk}_{\rm sup} \geq q_+
\right]=\lim_{q_-\rightarrow \infty}\mathsf{P}_+=  e^{-\beta q_+
}/\langle e^{\beta \Delta_+} \rangle_{+}\leq e^{-\beta q_+ }.\label{excatjx}
\end{eqnarray}  
The last inequality follows from $\Delta_+>0$ and thus $\langle e^{\beta
\Delta_+} \rangle_{+}>1$.    For the average, we have 
\begin{eqnarray}
\langle Q^{\rm hk}_{\rm sup} \rangle \leq k_B T. 
\label{excatjx2}
\end{eqnarray}   
Note that  now we have used supremum value $Q^{\rm
hk}_{\rm sup}$ instead of the maximum value $Q^{\rm hk}_{\rm max}$, since the maximum value  may not  exist
 for  processes that may have jumps.
 The tightness of the bounds~(\ref{excatjx}-\ref{excatjx2})  depend on the
magnitude of the overshoot variable $\langle e^{\beta \Delta_+}
\rangle_{+}$, which  is process dependent.  For continuous processes, $\Delta_+ = 0$, and therefore the statistics of the long-time supremum are universal.

%
%
%
The martingality of $e^{\beta Q^{\rm hk}_t}$ implies  that several  fluctuation properties of the housekeeping heat of continuous processes that do no satisfy detailed balance are universal.  These universal properties characterize  fluctuations that oppose the general tendency of the process to dissipate.   We have also found  bounds that hold for any isothermal Markovian nonequilibrium process.   These bounds are   stronger than what we would expect from the integral fluctuation relation.    In the following, we illustrate the bound~\eqref{boundhousex} for the cumulative distribution of the finite-time maximum value of the housekeeping heat over a finite time.

\begin{figure}[ht!]
\centering
\includegraphics[width=.36\textwidth]{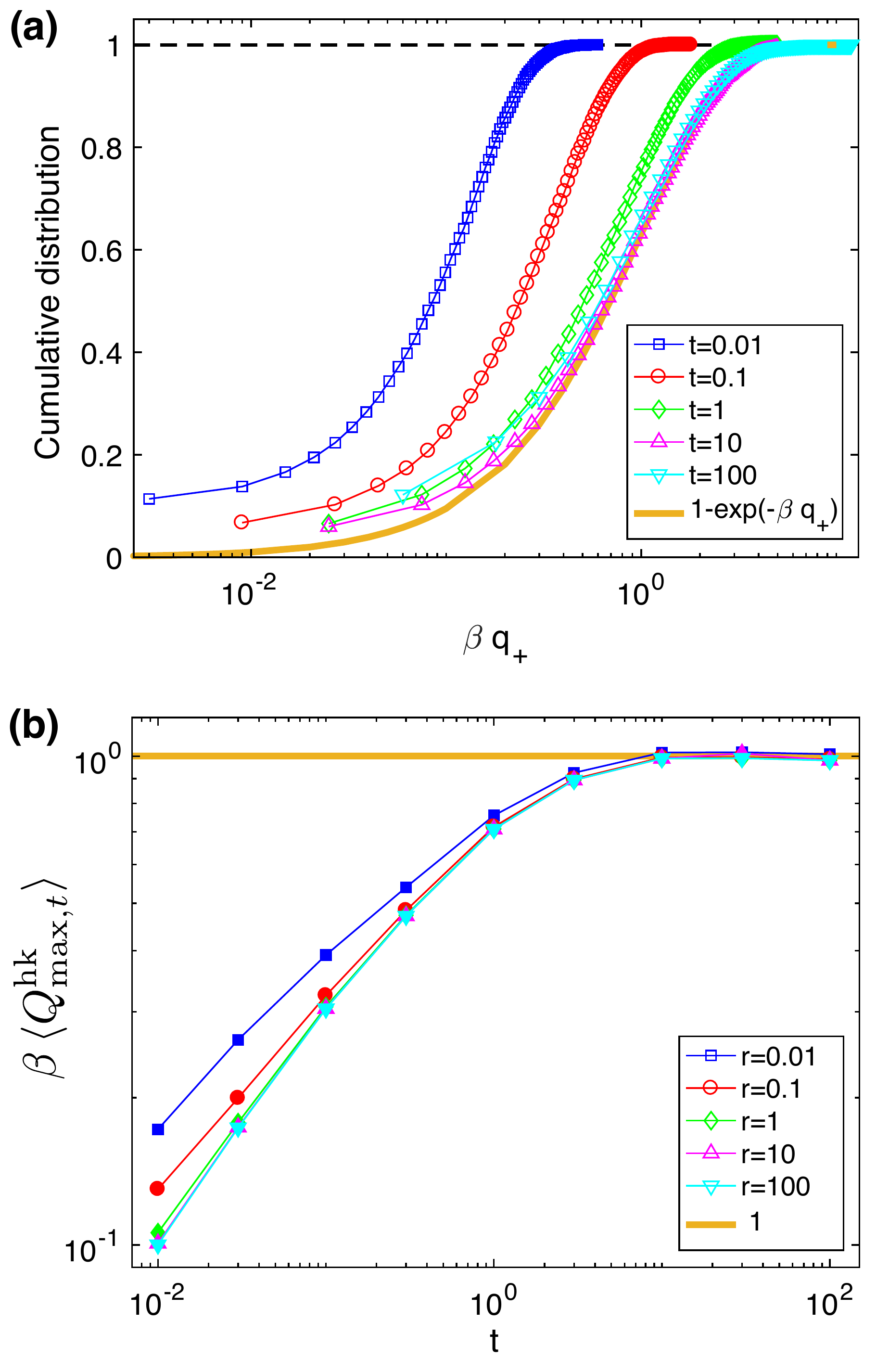}
\caption{Extreme-value statistics of the housekeeping heat. (a)
Cumulative distribution of the maximum value (symbols with lines) of the housekeeping heat
$Q^{\rm hk}_{\text{max},t}$  up to time $t$, for $r=1$ and different values of $t$ (legend). The yellow
line is the theoretical bound~(\ref{eq:lastnew}) and the black dashed line is a guide to the eye.  (b) Average value
$\langle Q^{\rm hk}_{\text{max},t}\rangle$ in units of $k_{\rm B}T$ as a
function of $t$ for different values of the driving speed $r$ (legend).
The yellow line is set to unity, corresponding to the bound (\ref{eq:lastnew2}). Here, we have taken $D=1$,
$\omega=1$, $10^4$ independent realizations of the dynamics with initial distribution $\rho_0(\theta)=\delta(\theta)$, and integration time step~$\Delta t = 10^{-4}$.
 \label{fig:models}}
\end{figure}

\section{Simulations} We now test our results for the statistics of the maximum of the housekeeping heat with numerical simulations of a model described by the following one-dimensional overdamped Langevin equation
\begin{align}
 \frac{\text{d}\theta_t}{\text{d}t} & =\omega-\partial_\theta
 V(\theta_t;t)+\sqrt{2D}~\eta_t.
 \label{eq:m1}
\end{align}
Here  $\eta_t$  is a  Gaussian white noise satisfying
$\langle\eta_t\rangle=0,\thinspace\langle\eta_t\eta_{t'}\rangle=\delta(t-t')$,
$D=1/\beta$ due to the fluctuation-dissipation relation (here we take
unity for the friction constant). The system is driven out of
equilibrium by a constant external torque $\omega$ and a time-dependent potential
\begin{align}
V(\theta;t) & =(1/\beta)\ln(\cos\theta+\lambda_{t});\thinspace\thinspace\lambda_{t}=2+rt,
 \label{eq:m2}
\end{align}
where $r>0$ is the speed of the time-dependent protocol $\lambda_t$. The corresponding   accompanying density is
\begin{align}
\pi_{t}(\theta)  =\mathcal{N}^{-1} e^{-\beta U(\theta)}\int_{\theta}^{\theta+2\pi}\text{d}\theta'\thinspace e^{\beta U(\theta')} ,
\label{eq:accompshamik}
\end{align}
where $U(\theta)\equiv-\omega\theta+V(\theta)$, and $\mathcal{N}$ is a
normalization constant~\cite{Reimann}. Using (\ref{eq:accompshamik})
in (\ref{eq:qhk}), we derive  the following analytical expression for
the housekeeping heat: 
\begin{equation}
\beta Q^{{\rm hk}}_{ t}
=-\int_{0}^{t}\frac{2\lambda_{s}+2\cos\theta_{s}}{2\lambda_{s}-\sin\theta_{s}-\cos\theta_{s}}\circ \text{d}\theta_{s}.
\label{eq:housekeepingmodel}
\end{equation}

We perform numerical simulations of Eq.~(\ref{eq:m1}) using Heun's
numerical integration scheme. his numerical scheme is shown to converge to the solution of Langevin equations of the type (\ref{eq:m1}) with the force interpreted in the Stratonovich sense \cite{Rumelin}, as  required to minimize numerical errors in stochastic energetics (see Sec. 4.1.2.5 in\,\cite{SekimotoBook}). From dynamical trajectories, we evaluate the housekeeping heat using (\ref{eq:housekeepingmodel}).   Trajectories of the fluctuating housekeeping heat  exhibit positive excursions corresponding to transient absorptions of housekeeping heat while on average tend to decrease in time (see Fig.~\ref{fig:traj}). From these trajectories, we then compute empirical cumulative distributions of the finite-time maximum of the housekeeping heat  $Q^{\rm hk}_t$ for a given value of $r$ (see Fig.~\ref{fig:models}a). The empirical distributions obey  the inequality $\text{Pr}\left[Q^{\rm hk}_{{\rm max},t} \leq q_+ \right] \geq 1- e^{-\beta q_+}$ for both small and large values of $t$. Interestingly, the inequality becomes tighter when $t$ is large, in agreement with (\ref{excatj}). Furthermore, we measure the average finite-time maximum of the housekeeping heat $\langle Q^{\rm hk}_{{\rm max},t}\rangle$ for different values of the driving speed $r$ and the observation time $t$. Our numerical results fulfil the inequality $\beta\langle Q^{\rm hk}_{{\rm max},t}\rangle\leq 1$ for all the tested parameter values, and the bound is tighter when $t$ is large, in agreement with our theory.

\section{Discussion} Our work demonstrates the power of martingale theory to describe extreme-value and stopping statistics   of the fluctuating housekeeping heat in nonequilibrium processes.  
Our results can be extended also to non-isothermal classical systems, for which  $e^{-S^{\rm a}_t/k_{\rm B}}$ is a path-probability ratio \cite{Esposito,VDB} and thus a martingale, i.e. $\langle e^{- S^{\rm a}_{t}/k_{\rm B}} \,|\, X_{[0,\tau]} \rangle  = e^{- S^{\rm a}_{\tau}/k_{\rm B}}$ for $t\geq\tau$. Here $S^{\rm a}_t$ is the \textit{adiabatic} entropy production i.e. the fluctuating 
entropy production required to keep the system out of equilibrium when the system is driven quasi-statically. Interestingly,  $S^{\rm a}_t$ follows the same fluctuation properties as $-Q^{\rm hk}_t/T$ concerning extrema and stopping-time statistics. We envisage that our theory could be tested experimentally with small colloidal, biological, electronic systems~\cite{Trapagnier,Toyabe,Jun,Martinez,Ciliberto,Ronzani} and that our results could also be extended to e.g. classical systems with long-range interactions~\cite{Shamik:16,Shamikbook} and quantum systems~\cite{Manzano}.

\textbf{Acknowledgments.-} The authors thank  MPIPKS, Dresden for
hosting them during the workshop ``Stochastic Thermodynamics: Experiment and Theory" in September 2018 when
major part of this work was done. We acknowledge fruitful discussions with Simone Pigolotti and Frank J\"ulicher.

\section{Appendix A: Proof of Eq.~\eqref{eq:15}}  The Ito expression for the housekeeping heat [cf. Eq.~\eqref{eq:10}] can be expressed in terms of the effective current $\vec{J}_t^{\pi}= \vec{F}_t \pi_t - \mathbf{D}_t\cdot \vec{\nabla}\pi_t$ as
\begin{eqnarray}
Q^{\rm hk}_t&=&- k_{\rm B}T \int_{0}^{t} \frac{ \mathbf{D}_{s}^{-1} \vec{J}^{\pi}_s}{\pi_s}(\vec{X}_{s})\cdot \mathrm{d}\vec{X}_{s} \nonumber\\
&-& k_{\rm B}T \int_{0}^{t}\text{d}s \left[  \mathbf{D}_s\cdot  \vec{\nabla}\cdot \left(\frac{ \mathbf{D}_{s}^{-1} \vec{J}^{\pi}_s}{\pi_s}\right) \right]\!(\vec{X}_s). \nonumber\\
\end{eqnarray} 
Using Ito's lemma in the above equation gives
\begin{eqnarray}
\text{d}e^{\beta Q^{\rm hk}_t} & =& -e^{\beta Q^{\rm hk}_t} \left[\frac{ \mathbf{D}_{t}^{-1} \vec{J}^{\pi}_t}{\pi_t}\right]\!(\vec{X}_{t})\cdot \mathrm{d}\vec{X}_{t} \nonumber\\
&-& e^{\beta Q^{\rm hk}_t}  \left[  \mathbf{D}_t\cdot  \vec{\nabla}\cdot \left( \frac{ \mathbf{D}_{t}^{-1} \vec{J}^{\pi}_t}{\pi_t} \right)  \right]\!(\vec{X}_t)  \text{d}t \nonumber\\
&+& \frac{1}{2}e^{\beta Q^{\rm hk}_t} \left[\frac{ \mathbf{D}_{t}^{-1} \vec{J}^{\pi}_t}{\pi_t}\right] \cdot 2\mathbf{D}_t \cdot\left[\frac{ \mathbf{D}_{t}^{-1} \vec{J}^{\pi}_t}{\pi_t}\right]\!(\vec{X}_t)\text{d}t. \nonumber\\\label{eq:29}
\end{eqnarray}
Substituting Eq.~(\ref{eq:langevin}) of the Main text in the first line
of~(\ref{eq:29}),  using the properties $\vec{\nabla}\cdot
\vec{J}^{\pi}_t = 0$ (which is a consequence of the
definition~(\ref{eq:Ac}) of the  accompanying distribution)  and the
mathematical property  $\nabla_\mu (\mathbf{D}^{-1}) =
-\mathbf{D}^{-1}\cdot(\nabla_\mu \mathbf{D})\cdot  \mathbf{D}^{-1} $
$-$with $\nabla_{\mu} $ given by the $\mu$ component of
$\vec{\nabla}$$-$, we obtain after some algebra 
\begin{eqnarray}
\frac{\mathrm{d}}{\mathrm{d}t} e^{\beta Q^{\rm hk}_t}= - \sqrt{2}  e^{\beta Q^{\rm hk}_t} \left[\frac{\vec{J}^{\pi}_t \cdot \mathbf{D}_t^{-1/2}}{\pi_t}\right]\!(\vec{X}_t) \cdot \vec{\xi}_t. \quad \square 
\end{eqnarray}


\newpage

\begin{thebibliography}{99}
\bibitem{Ville:39} J. Ville, \textit{Etude critique de la notion de collectif} (PhD Thesis)  Gauthier-Villars, Paris (1939).

\bibitem{Doob}  J. L. Doob, \textit{Stochastic processes} (Wiley, New York, 1953).

\bibitem{Williams} D. Williams, \textit{Probability with martingales}
(Cambridge University Press, Cambridge, 1991). 

\bibitem{LS} R. S. Liptser and A. N. Shiryaev,  \textit{Statistics of
random processes: I \& II}  (Springer Science \& Business Media, Berlin, 2013).

\bibitem{Oks} B. {\O}ksendal, \textit{Stochastic differential equations}
(Springer, Berlin, 2003).

\bibitem{Note} A sufficient condition for $I_t = \int_0^t F(X_{s})\cdot d\mathcal{W}_s$ to be a martingale is $\langle \int_0^t F^2(X_{s})ds\rangle  <\infty$ for all $t\geq 0$.
\bibitem{radon} Probability ratios of densities of trajectories are in fact Radon-Nikodym derivatives between two mutually absolute continuous measures. 


\bibitem{Pliska} S. Pliska, \textit{Introduction to mathematical
finance} (Blackwell publishers, Oxford, 1997).



\bibitem{Neri:17} I. Neri, \'E. Rold\'an and F. J\"ulicher, Phys. Rev. X \textbf{7}, 011019 (2017).

\bibitem{Chetrite:11} R. Chetrite and S. Gupta, J. Stat. Phys. \textbf{143}, 543 (2011).

\bibitem{Ventejou:18} B. Vent\'ejou and K. Sekimoto, Phys. Rev. E
\textbf{97}, 062150 (2018). 

\bibitem{bernard}M. Bauer and D. Bernard, Phys. Rev. A \textbf{84},
044103 (2011).

\bibitem{adler}S. L. Adler, D. C. Brody, T. A, Brun and L. F. Hughston,
J. Phys. \textbf{34}, 8795 (2001).

\bibitem{Singh:17} S. Singh \textit{et al.}, arXiv:1712.01693 (2017).

\bibitem{Oono:98} Y. Oono and M. Paniconi,  \text{Pr}og.  Theor. Phys. Suppl.  \textbf{130}, 29 (1998).

\bibitem{Seki} K. Sekimoto, Prog. Theor. Phys. Suppl. \textbf{130}, 17 (1998).

\bibitem{Sasa} T. Hatano and S.-i. Sasa, Phys. Rev. Lett. \textbf{86}, 3463 (2001).

\bibitem{Speck:05} T. Speck and U. Seifert, J. Phys. A, \textbf{38}, L581 (2005).

\bibitem{Esposito} M. Esposito and C. Van den Broeck, Phys. Rev. E,
\textbf{82}, 011143 (2010). 

\bibitem{VDB} C. Van den Broeck and M. Esposito, Phys. Rev. E
\textbf{82}, 011144 (2010). 

\bibitem{bounded} A \textit{bounded} stopping time $\mathcal{T}\in [0,\tau]$ with $\tau$ a constant fixed time. 

\bibitem{Helvetica} P. H\"anggi, Helv. Phys. Acta \textbf{53}, 491 (1980).

\bibitem{Ku2} Y. L. Klimontovich, Phys.-Usp. \textbf{37}, 737 (1994).

\bibitem{Ku1} Y. L. Klimontovich, Phys. A \textbf{163}, 515 (1990).

\bibitem{fisk} D. L. Fisk, \textit{Quasi-martingales and stochastic integrals}, Tech Rep 1 Dept Math Univ Michigan State (1963).

\bibitem{stratonovich} R. L. Stratonovich, \textit{Conditional Markov
process and their applications to the theory of optimal control} (Amer.
Elsiever, New York, 1968).

\bibitem{hanggi} P. H\"anggi, H. Thomas, Phys. Rep. \textbf{88}, 207 (1982).

\bibitem{Harris} R. J. Harris and G. M. Sch\"utz,  J. Stat. Mech. \textbf{2007}(07), P07020 (2007).

\bibitem{CMP} R. Chetrite and K. Gawedzki, Comm. Math. Phys.
\textbf{282}(2), 469 (2008).

\bibitem{Reinaldo:10} R. Garc\'ia-Garc\'ia, D. Dom\'inguez, V. Lecomte
and A. B. Kolton, Phys. Rev. E \textbf{82}, 030104(R) (2010). 

\bibitem{esposito07}M. Esposito, U. Harbola  and S. Mukamel, Phys.
Rev. E \textbf{76}, 031132 (2007).

\bibitem{Simone} S. Pigolotti, I. Neri, \'E. Rold\'an and F. J\"ulicher, Phys. Rev. Lett. \textbf{119}, 140604 (2017).

\bibitem{Heston} S. L. Heston, Rev. Financial Stud. \textbf{6}(2), 327-343 (1993).

\bibitem{Chun}  H.-M. Chun  and J. D. Noh, arXiv:1810.01121 (2018).


\bibitem{Reimann} P. Reimann, Phys. Rep., \textbf{361}, 57 (2002).

\bibitem{Rumelin} W. R\"umelin,  SIAM J. Num. Anal., \textbf{19}(3), 604-613 (1982).

\bibitem{SekimotoBook} K. Sekimoto, \textit{Stochastic Energetics} (Springer, Berlin, 2010). 

\bibitem{Trapagnier} E. H. Trepagnier, C. Jarzynski,  F. Ritort, G. E. Crooks, C. J. Bustamante, and J. Liphardt, PNAS, \textbf{101}(42), 15038-15041 (2004). 

\bibitem{Toyabe}  S. Toyabe, T. Watanabe-Nakayama, T. Okamoto, S. Kudo, and E. Muneyuki, PNAS, \textbf{108}(44), 17951-17956 (2011).

\bibitem{Jun} Y. Jun, M. Gavrilov, and J. Bechhoefer, Phys. Rev. Lett., \textbf{113}(19), 190601 (2014).

\bibitem{Martinez} I. A. Mart\'inez, \'E. Rold\'an, L. Dinis, and R. A. Rica, Soft Matter, \textbf{13}(1), 22-36 (2017).

\bibitem{Ciliberto} S. Ciliberto, Phys. Rev. X, \textbf{7}(2), 021051 (2017). 

\bibitem{Ronzani} A. Ronzani, B. Karimi, J. Senior, Y.-C. Chang, J. T. Peltonen, C. Chen and J. P. Pekola,  Nature Phys. \textbf{14}, 991-995 (2018).

\bibitem{Shamik:16} S. Gupta, T. Dauxois and S. Ruffo, EPL \textbf{113}, 60008 (2016).

\bibitem{Shamikbook} S. Gupta, A. Campa and S. Ruffo, \textit{Statistical Physics of Synchronization} (Springer, Berlin, 2018).


\bibitem{Manzano} G. Manzano, J. M. Horowitz, and J. M. R. Parrondo, Phys. Rev. X, \textbf{8}(3), 031037 (2018).


\end{thebibliography}
\end{document}